\documentclass[aps, prl, preprint, showpacs,preprintnumbers, a4paper]{revtex4-1}
\usepackage[utf8x]{inputenc}
\usepackage[english]{babel}
\usepackage{graphicx}
\usepackage{dcolumn}
\usepackage{amsmath}
\usepackage{amssymb}
\usepackage{bm}
\usepackage{color}

\begin{document}
\title{Instabilities of localized structures in dissipative systems with delayed feedback}
\author{S. V.~Gurevich}
\email{gurevics@uni-muenster.de}
\author{R.~Friedrich}
\affiliation{Institute for Theoretical Physics, University of M\"unster,
Wilhelm-Klemm-Str.\,9, D-48149 M\"unster, Germany\\
}
\date{\today}
\begin{abstract}
We report on a novel behavior of  solitary localized
structures in a real Swift-Hohenberg equation subjected to a delayed feedback.
We shall show that variation in the product of the delay time and the feedback
strength leads to nontrivial instabilities resulting in the formation of oscillons, soliton rings, 
labyrinth patterns or moving structures. We provide a bifurcation analysis of the delayed system 
and derive a system of order parameter equations explicitly describing the temporal behavior of the localized structure
in the vicinity of the bifurcation point. We demonstrate that a normal form of the bifurcation, responsible for the emergence of moving solitary structures can be obtained and show that spontaneous motion to the lowest order occurs without change of the shape.  
\end{abstract}
%
%
\maketitle

Control and engineering of complex spatio-temporal patterns in high-dimensional non-equilibrium systems has evolved as one of the central issues in applied nonlinear science~\cite{Mikhailov2006, SchoellChaos2009}. Among other control techniques, a quite simple and efficient scheme is the time delayed feedback (TDF) ~\cite{PyragasPRA1992}. Although TDF method was originally designed to control dynamical systems with a few degrees of freedom, it has been successfully applied to a large number of theoretical and experimental high-dimensional spatially extended systems~\cite{PierrePRL1996, *LuethjePRL2001, *BabaPRL2002, *SchlesnerPRE2003, *Postlethwaite200765, *Kyrychko2009}. In particular, control of dissipative solitary localized structures (LSs) has been of increasing interest in recent years~\cite{AkhmedievDS2008, PurwinsDS2010}. Among other LSs, traveling pulses and fronts subjected to a TDF appear in various contexts~\cite{DahlemCahos2008,*Schneider2009, *Dahlem2010, *FortPRL2002, *
OrtegaCejas2004, *Erneux2010}. Recently, properties of two-dimensional LSs by means of a real Swift-Hohenberg equation subjected to TDF were studied in~\cite{TlidiPRL2009}. It was shown that when the product of the delay time and the feedback strength exceeds some critical value, a single LS starts to move in an arbitrary direction. Moreover, an analytical formula for its velocity was derived. We refer to such kind of instability as \emph{delay-induced drift-bifurcation}.

In this Letter we investigate the stability of a single LS in a real Swift-Hohenberg equation subjected to a TDF  in details. We show that varying of the product of the delay time and the feedback strength, apart from moving LSs, also leads to the formation of complex spatio-temporal patterns, such as oscillons, soliton rings or labyrinths, which are not generic for the Swift-Hohenberg equation without TDF. We provide a bifurcation analysis of the delayed system and derive a set of order parameter equations for the position of the LS as well as for its shape. In addition, a normal form normal form of the delay-induced drift-bifurcation is derived, showing that the delay-induced motion can arise without pronounced change of the LS's shape. The bifurcation analysis is performed in general form and therefore can be applied to any system possessing solutions in form of LS's. 

We start with the delayed Swift-Hohenberg equation (DSHE) in the form
\begin{equation}\label{eq:DSHE}
  \frac{\partial q_t}{\partial t}=-a_1\,\nabla^2\,q_t-a_2\,\nabla^4\,q_t+f(q_t)+\alpha\,\bigl(q_t-q_{\,t-\tau}\bigr)\,,
 \end{equation}
where $q_t:=q(\mathbf{x},\,t)$, $\mathbf{x}=(x,\,y)^T$ is the real distributed order parameter, $a_1,\, a_2 >0$, $\tau$ is the delay time, whereas $\alpha$ denotes the positive delay strength. The nonlinear function $f(q_t)=Y+C\,q_t-q_t^3$ is a cubic polynomial with $C>0$, while the sign of $Y$ is arbitrary. For $\alpha=0$ Eq.~\eqref{eq:DSHE} becomes the Swift-Hohenberg equation (SHE), which often serves as a paradigm for general pattern forming system and arises in a number of applications including fluid dynamics, nonlinear optics and other fields~\cite{CrossHohenberg,*TlidiPRL1994, *RichterPRL2005}. In the context of nonlinear optics, the Eq.~\eqref{eq:DSHE} describe, e.g., a spatial-temporal dynamics in a
coherently driven optical resonator subjected to TDF~\cite{TlidiPRL2009}. Here $q_t$ and  $Y$ are the deviation of the electric and injection fields, respectively,  and $C$ is the cooperative parameter. The derivation of the Eq.~\eqref{eq:DSHE} as well as detailed description of physical meaning of the parameters can be found in~\cite{TlidiPRL2009, TlidiEPJD2010}. Note that the SHE possesses a Lyapunov functional, i.e., any perturbation evolves towards its local minima. That is, no Hopf bifurcations or traveling waves are possible. Furthermore, stable stationary solutions of the SHE occupy these minima and satisfy $\partial q_t/\partial t=0$. However, in the presence of TDF Eq.~\eqref{eq:DSHE} loses the gradient structure. Although the stationary solutions are not affected by the TDF, their stability properties may change.

\noindent We are interested in the stability of the stationary localized solution of Eq.~\eqref{eq:DSHE}. In the simplest case it is a single spot-like structure with rotational symmetry. Notice that more complex LSs like rings, localized hexagon patches or stripes can also be found~\cite{LloydSIAM2008}, but they are out of scope of this paper. 	 
The function $q_t(\mathbf{x},\,t)$, given by Eq.~\eqref{eq:DSHE} is a scalar quantity. Nevertheless for the following stability and bifurcation analysis it is convenient to consider Eq.~\eqref{eq:DSHE} in terms of a $n$-dimensional vector function $\mathbf{q}=\mathbf{q}(\mathbf{x},t)=\{ q_{ti}\}$, $\mathbf{x}=(x,\,y)^T$
\begin{equation}\label{eq:GenEq}
\partial_t\mathbf{q}(\mathbf{x},t)=\mathfrak{L}\,\mathbf{q}(\mathbf{x},t)+\alpha\,\mathrm{E}\,\bigl(\mathbf{q}(\mathbf{x},t)-\mathbf{q}(\mathbf{x},t-\tau)\bigr)\,,
\end{equation}
keeping in mind the scalar Eq.~\eqref{eq:DSHE} for the numerics. Here, $\mathfrak{L}$ denotes the nonlinear Swift-Hohenberg operator and $\mathrm{E}$ is the coupling matrix. It describes the particular coupling of the feedback term and is given by an identity matrix for simplicity of the further analysis. Assume that the stationary localized spot solution of~\eqref{eq:GenEq} $\mathbf{q_0}(\mathbf {x})$ exists. For $\alpha=0$ linear stability analysis of $\mathbf{q_0}$ yields a real spectrum, given by the linear eigenvalue problem $\mathfrak{L}'(\mathbf{q_0})\,\bm{\varphi}=\mu\,\bm{\varphi}$, since the linearization operator $\mathfrak{L}'(\mathbf{q_0})$ is self-adjoint. Due to translational invariance $\mu=0$ is an eigenvalue of $\mathfrak{L}'(\mathbf{q_0})$, corresponding to two independent neutral eigenfunctions $\bm{\varphi}_{\mathbf{x}}^{\mathcal{G}}(\mathbf{x})$, also referred to as Goldstone modes. 
The discrete spectrum of $\mathfrak{L}'(\mathbf{q_0})$ is usually located near zero and is well separated from the rest of the spectrum. It contains modes $\propto\bm{\varphi}_n(\mathbf{x})\,e^{in\phi}$ with $n=0,\,\pm2$, where the mode with $n=0$ results in the change of the size of the LS, and $n=\pm 2$ lead to deformations. We suppose that for $\alpha=0$ the stationary solution $\mathbf{q_0}$ is stable, i.e.,  all eigenvalues but $\mu=0$ have negative real parts. For $\alpha>0$ the spectrum of the corresponding linear operator can be constructed from the spectrum of $\mathfrak{L}'(\mathbf{q_0})$, since the latter commutes with the identity coupling matrix. Indeed, both operators  possess the same set of eigenfunctions $\bm{\varphi}(\mathbf{x})$, whereas the complex eigenvalues $\lambda$ of the delayed problem are determined as solutions of the transcendental equation~\cite{TlidiPRL2009}: 
\begin{equation}\label{eq:LinStab}
 \mu=\lambda-\alpha\,\bigl(1-e^{-\lambda\,\tau}\bigr)\,,
\end{equation}
that can be found in terms of Lambert $\mathrm{W}$-functions~\cite{Corless96} as $\lambda=\mu+\alpha+\frac{1}{\tau}\,\mathrm{W}_m\bigl(-\alpha\tau\,\exp\bigl(-(\mu+\alpha)\tau\bigr)\bigr)\,,\quad m\in\mathbb{Z}$. Hence, the stationary solution of~\eqref{eq:GenEq} remains stable, if $\mathrm{Re}(\lambda(\mu))<0$ for any $\mu$ from the spectrum of  $\mathfrak{L}'(\mathbf{q_0})$. Bifurcation points correspond to $\mu$, at which $\mathrm{Re}(\lambda(\mu))$ vanishes. In particular, instability criterion for all eigenvalues $\lambda$, corresponding to the real eigenvalues $\mu$, is given by the relation $a:=\alpha\tau\geq \max\bigl\{-\mu\,\tau/2\,,1\bigr\}\,$. One can see that $\mu=0$ yields the bifurcation point $a=1$, corresponding to the onset of spontaneous motion, observed in~\cite{TlidiPRL2009}. Notice that eigenvalues $\mu$ of discrete critical modes $n=0,\, n=\pm2$  are close to zero, i.e., for typical system parameters the instability threshold yields the value $a=1$, which becomes the 
bifurcation point of higher codimension. The remaining part of the spectrum is well separated from zero and is not responsible for primary instabilities. In addition, using $a$ as a control parameter, one can find a critical delay time $\tau_c:=x\,a/\mu$ for which each of the modes becomes unstable by means of the solvability condition:
\begin{equation}\label{eq:SolvCond}
 \pm \mathrm{arccos}(1+x)+2\pi m=\pm a\,\sqrt{1-(1+x)^2},\,\quad m\in\mathbb{Z}\,.
\end{equation}

\begin{figure}
  \includegraphics[width=0.35\textwidth]{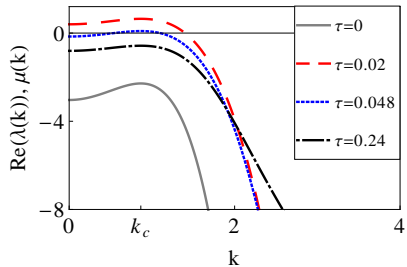}
 \caption{\label{FigStabHom} A real part of the dispersion function $\lambda(k)$, calculated at $a=1.05$ for three different values of the delay time $\tau$, corresponding to a homogeneous Hopf bifurcation for $\tau=0.02$ (dashed rot line), a traveling wave bifurcation for $\tau=0.048$ (dotted blue line) as well as stable solution  $\mathbf{q_h}$ for $\tau=0.24$ (black dash-dotted line). Only the main branch of $\mathrm{Re}(\lambda(k))$ is shown. A gray solid line corresponds to the function $\mu(k)$ for $\tau=0$.}
 \end{figure}

\noindent The next point to emphasize is that the TDF can also affect the stability of the spatial uniform solution $\mathbf{q_h}$ of the system~\eqref{eq:GenEq}~\cite{TlidiEPJD2010}. Suppose that $\mathbf{q_h}$ is stable for $\tau=0$, i.e., the dispersion function $\mu(k)=a_1k^2-a_2k^4+f'(\mathbf{q_h})<0$ for all values of $k$. For $\tau>0$ the linear stability analysis with respect to perturbations  $\propto\exp(i\,k\,\mathbf{x}+\lambda(k)t)$ yields the same Eq.~\eqref{eq:LinStab}. The dependence of $\mathrm{Re}(\lambda(k))$ on the delay time $\tau$, calculated for fixed value of $a=1.05$ is presented in Fig.~\ref{FigStabHom}. 
One can see that for $\tau=0$ $\mathbf{q_h}$ is stable. However, for $\tau>0$ the behavior of $\mathrm{Re}(\lambda(k))$ indicates the existence of a traveling wave bifurcation (TW) with a finite wave number $k=k_c=\pm\sqrt{a_1/2a_2}$~\cite{TlidiEPJD2010}. Here, for rather small values of $\tau$ the finite band of unstable wave-numbers includes $k=0$ (see Fig.~\ref{FigStabHom} for $\tau=0.02$), what corresponds to a so-called homogeneous Hopf bifurcation (HH). An increasing of $\tau$ leads to the decreasing of the width of the band and $\lambda(k=0)$ becomes negative for some $\tau$ (Fig.~\ref{FigStabHom} for $\tau=0.048$). Further increase in $\tau$ leads to the stabilization of  $\mathbf{q_h}$ as indicated in Fig. 1 for $\tau=0.24$. That is, the stability of $\mathbf{q_h}$ is violated only for small delay times. Notice that the behavior of $\mathrm{Re}(\lambda(k))$ for large $k$ is different for different values of $\tau$. While for small delays $\mathrm{Re}(\lambda(k))$ decays as $k^4$,  it decays rather as $k^2$ for larger $\tau$, which is more in the nature of e.g., reaction-diffusion systems. Hence, in this parameter region one would expect solutions which are not generic for the SHE.

\begin{figure}
  \includegraphics[width=0.30\textwidth]{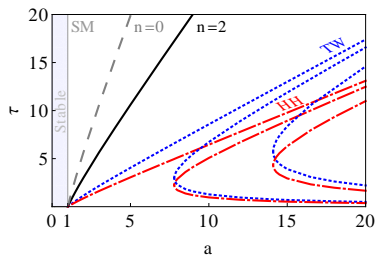}
  \caption{\label{StabDiag} Bifurcation diagram in $(a,\,\tau)$ plane. Different lines separate instability regions, 
  corresponding to different modes that are unstable below the corresponding line. 
  The main solution branch for $n=0$ (dashed gray line) and $n=2$ (solid black
line) as well as three branches for instability thresholds of traveling
wave (TW) bifurcation (dotted blue curves) and homogeneous Hopf (HH) bifurcation
(dashed-dotted red curves) are shown. The vertical line at $a=1$ indicated the
onset of spontaneous motion (SM).}
 \end{figure}

\begin{figure}
  \includegraphics[width=0.40\textwidth]{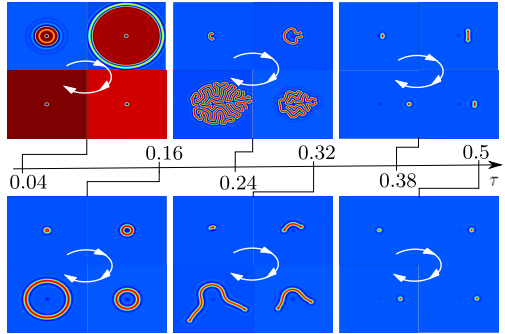}
 \caption{\label{FigDNS} Numerical solutions of Eq.~\eqref{eq:DSHE}, obtained for fixed $a=1.05$ and different values of the delay time $\tau$. Numerical simulations have been performed on the two-dimensional square domain $100\times 100$ using a pseudo-spectral method with $512\times 512$ grid points, whereas a Runge-Kutta 4 scheme is employed for the time stepping. Parameters are: $a_1=-2$, $a_2=-4/3$, $C=1$, $Y=-0.4$. White arrows indicate the time evolution direction.}
 \end{figure}

In Fig.~\ref{StabDiag} a bifurcation diagram in $(a,\,\tau)$ plane is presented. Each line corresponds to the solution of Eq.~\eqref{eq:SolvCond}, calculated for different $a$ and represents the instability thresholds for critical modes as well as thresholds for HH and TW bifurcations.
 One can see that for $a=1$ both solutions $\mathbf{q_0}$ and $\mathbf{q_h}$ become unstable. However, for $a>1$ the instability type depends on the value of $\tau$. This is illustrated in Fig.~\ref{FigDNS}, where results of numerical simulations of Eq.~\eqref{eq:DSHE} for $a=1.05$ and different values of $\tau$ are shown. 
 For small $\tau$ all modes are excited (see Fig.~\ref{FigDNS} for $\tau=0.04$) and a stationary bright LS transforms into a breathing dark LS. Here, the steady state transfers to a new oscillatory state of a high positive amplitude, i.e., a transition from the bright LS to a dark oscillon is observed.  If $\tau$ is increased, $\mathbf{q_h}$ becomes stable, but the spatial modes remain unstable and a soliton ring ($\tau=0.16$) or a labyrinth ($\tau=0.24$) is formed. Indeed, a LS moves with a rather large velocity, its shape changes ($n=0$) and deforms ($n=2$). Finally, the impact of the curvature instability results in the formation of the labyrinth pattern. Note that labyrinths are not solutions of the SHE. However, as mentioned above, there exist parameter regions, where the DSHE behaves like a reaction-diffusion system, where the formation of labyrinths is well understood~\cite{GoldSteinPRE1996}. This indicates that for certain values of $\tau$ the DSHE to some extend operates in the regime of reaction-diffusion systems. With increase in $\tau$ the amplitudes of unstable modes decrease and instead of  labyrinths, a moving wave segment emerges on the same time scale ($\tau=0.32$).  For $\tau=0.38$, both modes $n=0$ and $n=2$ are stable, i.e., solutions in the form of moving LSs are expected. However, close to the instability threshold the real parts of corresponding eigenvalues are rather small, what affects the LS's shape on a rather large time scale. Finally, a motion with a constant velocity is observed as shown in Fig.~\ref{FigDNS} for $\tau=0.5$. Here, the dynamics of the LS is little affected by critical modes and the LS propagates without pronounced change of its shape. 

\begin{figure}
 \begin{tabular}{ll}
 (a) & (b) \\
  \includegraphics[width=0.24\textwidth]{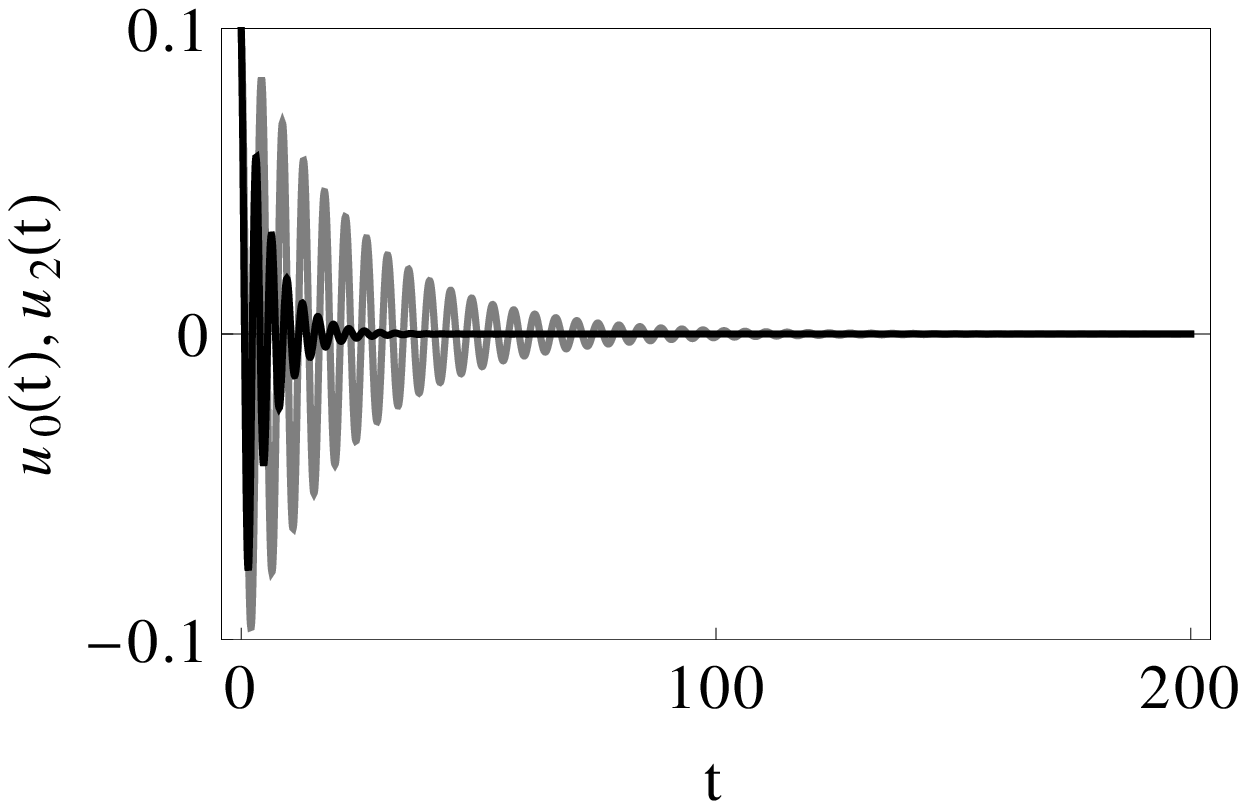}& \includegraphics[width=0.24\textwidth]{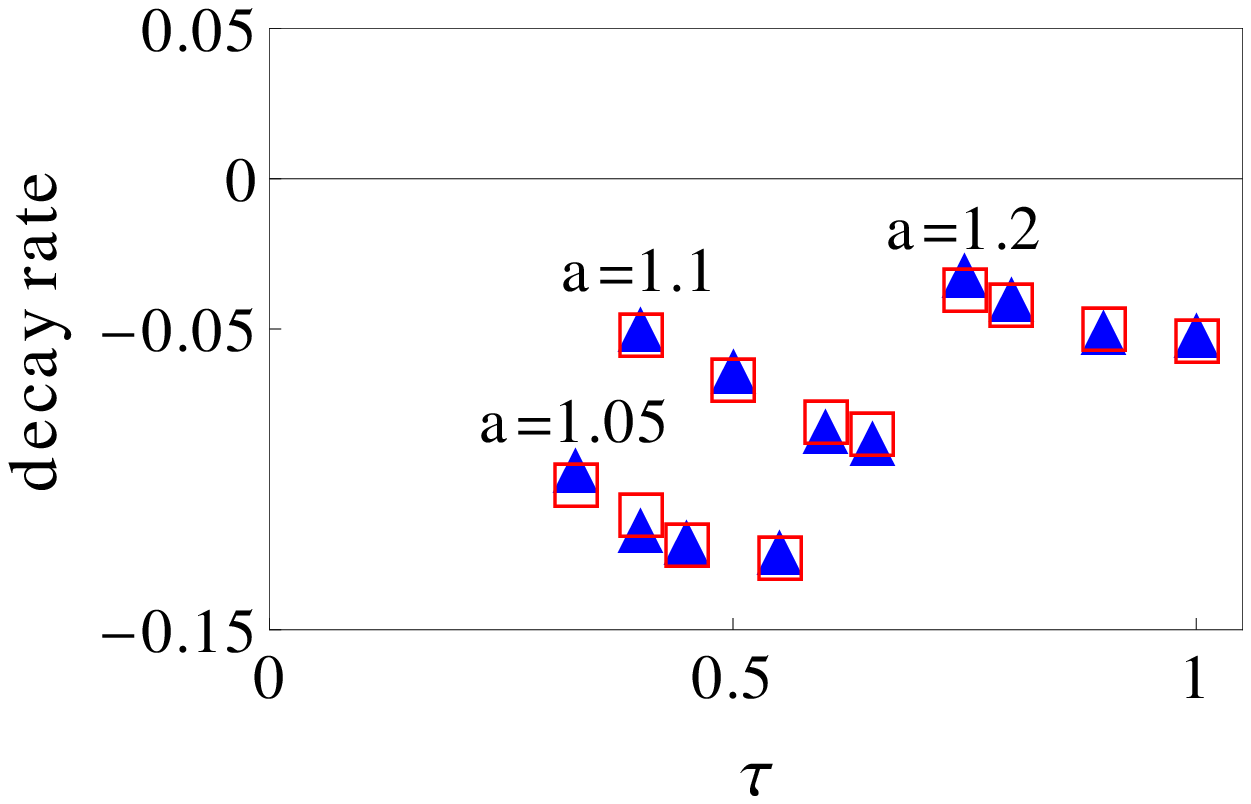}
 \end{tabular}
 \caption{\label{FigCompare} (a) Time evolution of the amplitudes $u_0(t)$ (gray) and $u_2(t)$ (black), as solutions of Eqs.~\eqref{eq:ShapeDelay}, calculated for $\tau=0.38$ and $a=1.05$. Other parameters are the same as in Fig.~\ref{FigDNS}.  (b) Dependence of the instability increment on $\tau$ for three different values of $a$ as result of reduce model~\eqref{eq:ShapeDelay}-\eqref{eq:ShiftDelay} (filled blue triangles) and numerical simulations of Eq.~\eqref{eq:DSHE}                       (open red squares).}
\end{figure}

In order to investigate the stability of the LS in the presence of the TDF from the point of view of bifurcation theory we perform the following ansatz
\begin{equation}\label{eq:Ansatz}
 \mathbf{q}(\mathbf{x},\,t)=e^{-\mathbf{R}(t)\cdot\nabla}\,\biggl(\mathbf{q_0}(\mathbf{x})+\mathbf{w}(\mathbf{x},\,t)\biggr)\,.
\end{equation}
Here, the operator $e^{-\mathbf{R}(t)\cdot\nabla}$ generates a shift of the pattern to the position, defined by the vector $\mathbf{R}(t)$ and
$\mathbf{w(\mathbf{x}, t)}=\sum_j\,\boldsymbol{\varphi}_j(\mathbf{x})\,e^{\lambda_j\,t}$ is the shape deformation, given by a sum of all stable modes of the delayed problem. Our goal is to decompose Eq.~\eqref{eq:GenEq} in such a way, that evolution equations for the shift $\mathbf{R}(t)$ as well as for the change of the shape in terms of $\mathbf{w}(\mathbf{x}, t)$ are derived. An equation for $\mathbf{R}(t)$ can be obtained by requiring that the shape deformation is orthogonal to the Goldstone mode $\bm{\varphi}_{\mathbf{x}}^{\mathcal{G}}$. This condition leads to
\begin{eqnarray}\label{eq:shift}
        -\dot{\mathbf{R}_t}\bigl(1-\langle\boldsymbol{\varphi}_{\mathbf{x}}^{\mathcal{G}}|\mathbf{w}_t\rangle\bigr)&&=-\langle\ \boldsymbol{\varphi}_{\mathbf{x}}^{\mathcal{G}}|\mathfrak{L}'(\mathbf{q}_0)\,\mathbf{w}_t\rangle-\langle\ \boldsymbol{\varphi}_{\mathbf{x}}^{\mathcal{G}}|\mathrm{\bf N}(\mathbf{w}_t)\rangle\\
        \nonumber &&-\alpha\langle\ \boldsymbol{\varphi}_{\mathbf{x}}^{\mathcal{G}}|\mathrm{\bf M}(\mathbf{x},\,t,\,\mathbf{R}_t-\mathbf{R}_{t-\tau},\,\mathbf{w}_t,\,\mathbf{w}_{t-\tau})\rangle\,.
   \end{eqnarray}
   Here, we introduce the nomenclature $\mathbf{R}_t:=\mathbf{R}(t)$, $\mathbf{R}_{t-\tau}:=\mathbf{R}(t-\tau)$, $\mathbf{w}_t:=\mathbf{w}(\mathbf{x},t)$ and $\mathbf{w}_{t-\tau}:=\mathbf{w}(\mathbf{x},t-\tau)$, the vector  
   $$ 
   \mathrm{\bf N}(\mathbf{w}_t)=\frac{1}{2!}\,\mathfrak{L}''(\mathbf{q}_0)\mathbf{w}_t\mathbf{w}_t+\frac{1}{3!}\,\mathfrak{L}'''(\mathbf{q}_0)\mathbf{w}_t\mathbf{w}_t\mathbf{w}_t
   $$ 
   contains nonlinear contributions, whereas the delay terms are lumped into the vector 
   \begin{eqnarray*}
   \mathrm{\bf M}&=&\mathrm{E}\,\left(1-e^{(\mathbf{R}_t-\mathbf{R}_{t-\tau})\cdot\nabla}\right)\,\left(\mathbf{q}_0(\mathbf{x})+\mathbf{w}(\mathbf{x},t)\right)\\
   &&+\mathrm{E}\,e^{(\mathbf{R}_t-\mathbf{R}_{t-\tau})\cdot\nabla}\,\left(\mathbf{w}_t-\mathbf{w}_{t-\tau}\right)\,.
  \end{eqnarray*}
 The evolution equation for $\mathbf{w}(\mathbf{x}, t)$ is obtained by inserting Eq.~\eqref{eq:shift} into Eq.~\eqref{eq:GenEq}:
\begin{equation}\label{eq:shape}
   \frac{\partial \mathbf{w}_t}{\partial t}=\widetilde{\mathfrak{L}}'\mathbf{w}_t+\dot{\mathbf{R}}\frac{\widetilde{\partial}\mathbf{w}_t}{\partial\mathbf{x}}+\widetilde{\mathrm{\bf N}}+\alpha\widetilde{\mathrm{\bf M}}\,.
  \end{equation}
Here, $\widetilde{\mathfrak{L}}'\mathbf{w}_t=\mathfrak{L}'\mathbf{w}_t-\boldsymbol{\varphi}_{\mathbf{x}}^{\mathcal{G}}\,\langle\ \boldsymbol{\varphi}_{\mathbf{x}}^{\mathcal{G}}|\mathfrak{L}'\mathbf{w}_t\rangle/\langle\left(\boldsymbol{\varphi}_{\mathbf{x}}^{\mathcal{G}}\right)^2\rangle$ and other tilded terms are defined according to the same rule. Equations~\eqref{eq:shift}--\eqref{eq:shape} show that the deformation of the LS is given by a delayed equation and can be treated by a suitable bifurcation analysis. To this end, the form of the shape deformation should be specified. As a first approximation, we can perform a mode truncation, representing  $\mathbf{w}(\mathbf{x}, t)$ as a superposition of the modes of the discrete spectrum 
   $$
   \mathbf{w}(\mathbf{x},\,t)=u_0(t)\boldsymbol{\varphi}_0(\mathbf{x})+u_2(t)\boldsymbol{\varphi}_2(\mathbf{x})e^{i2\phi}+c.c.\,,
   $$
   where the slow functions of time $u_0(t)$ and $u_2(t)$ are the amplitudes of modes $\boldsymbol{\varphi}_0(\mathbf{x})$ and $\boldsymbol{\varphi}_2(\mathbf{x})$. Projection of Eq.~\eqref{eq:shape} onto these modes leads to a set of delayed equations:
\begin{eqnarray}\label{eq:ShapeDelay}
   \nonumber \dot{u}_0&=&\mu_0\,u_0+a_0\,u_0^2+b_0\,|u_2|^2+c_0\,u_0^3+d_0\,u_0|u_2|^2\\ 
   &&+\alpha(u_{0,t}-u_{0,t-\tau})\,,\\
    \nonumber \dot{u}_2&=&\mu_2\,u_2+b_2\,u_0u_2+c_2\,u_0^2u_2+d_2\,|u_2|^2u_2\\ 
    &&+\alpha(u_{2,t}-u_{2,t-\tau})\,. \nonumber
  \end{eqnarray}
 Here $\mu_0$ and $\mu_2$ are the eigenvalues of $\mathfrak{L}'(\mathbf{q_0})$, corresponding to the modes $n=0$, $n=2$, and all coefficients are determined by the integrals $a_0=\langle\boldsymbol{\varphi}_0|\mathfrak{L}''\boldsymbol{\varphi}_0^2\rangle/2$, $b_0=\langle\boldsymbol{\varphi}_0|\mathfrak{L}''\boldsymbol{\varphi}_2^2\rangle$, $c_0=\langle\boldsymbol{\varphi}_0|\mathfrak{L}'''\boldsymbol{\varphi}_0^3\rangle/6$, $d_0=\langle\boldsymbol{\varphi}_0|\mathfrak{L}'''\boldsymbol{\varphi}_0\boldsymbol{\varphi}_2^2\rangle$, $b_2=\langle\boldsymbol{\varphi}_2|\mathfrak{L}''\boldsymbol{\varphi}_0\boldsymbol{\varphi}_2\rangle$, $c_2=\langle\boldsymbol{\varphi}_0|\mathfrak{L}'''\boldsymbol{\varphi}_0^2\boldsymbol{\varphi}_2\rangle/2$, $d_2=\langle\boldsymbol{\varphi}_2|\mathfrak{L}'''\boldsymbol{\varphi}_2^3\rangle/2$. Moreover, the delayed equation for $\mathbf{R}(t)$ reads
  \begin{eqnarray}\label{eq:ShiftDelay}
  \nonumber
  \dot{\mathbf{R}_t}&=&\alpha\,\biggl(1-\sum_{k=0,\pm2}D^k(u_{k,t}-u_{k,t-\tau})\biggr)\,\bigl(\mathbf{R}_t-\mathbf{R}_{t-\tau}\bigr)\\ 
     &&-\beta\,\bigl|\mathbf{R}_t-\mathbf{R}_{t-\tau}\bigr|^2\bigl(\mathbf{R}_t-\mathbf{R}_{t-\tau}\bigr)\,,
    \end{eqnarray}
    with $\beta=\langle\left(\nabla\boldsymbol{\varphi}_{\mathbf{x}}^{\mathcal{G}}\right)^2\rangle/\langle\left(\boldsymbol{\varphi}_{\mathbf{x}}^{\mathcal{G}}\right)^2\rangle$. Thereby, the elements of the matrix $D^k$ are defined as $D_{ij}^k=\langle\boldsymbol{\varphi}_{i}^{\mathcal{G}}|\partial \boldsymbol{\varphi}_k(\mathbf{x})/\partial x_j\rangle\,.$ A set of nonlinear delayed differential equations~\eqref{eq:ShapeDelay}--\eqref{eq:ShiftDelay} is a system of \emph{order parameter equations}, describing the dynamical behavior of a single LS in the vicinity of the bifurcation point $a=1$. Numerical solution of ~\eqref{eq:ShapeDelay}--\eqref{eq:ShiftDelay} gives a time evolution of the amplitudes of the modes $u_0(t)$ and $u_2(t)$ as well as the evolution of the position $\mathbf{R}(t)$. In Fig.~\ref{FigCompare}~(a) numerical solution of Eq.~\eqref{eq:ShapeDelay}, showing  relaxation oscillations of both critical modes is presented (compare with Fig.~\ref{FigDNS} for $\tau=0.38$). Note that Eqs.~\eqref{eq:ShapeDelay} admit solutions with $u_2=0$, represented by interplay between the stable decaying breathing mode $n=0$ and spontaneous motion. Solving the system ~\eqref{eq:ShapeDelay}--\eqref{eq:ShiftDelay} with $u_2=0$ numerically, an instability increment, calculated from the time evolution of the amplitude $u_0(t)$, can be derived and compared with a decay strength, obtained from numerical simulations of Eq.~\eqref{eq:DSHE} (see Fig.~\ref{FigCompare}~(b)).
Note that Eqs. \eqref{eq:ShiftDelay}-\eqref{eq:ShapeDelay} are delayed differential equations with still infinitely many degrees of freedom. However they can be reduced to a set of ordinary differential equations for the solutions with $u_0=0$, corresponding to the case of a spontaneous motion. Let us define the drift velocity $\mathbf{V}(t):=\dot{\mathbf{R}}(t)$. Close to the bifurcation point $a=1$ the velocity is slowly varying quantity, i.e  $\mathbf{R}(t-\tau)=\mathbf{R}(t)-\mathbf{V}(t)\,\tau+\tau^2\dot{\mathbf{V}(t)}(t)/2+\mathcal{O}\left(d^2\mathbf{V}/dt^2\right)$. The result is \emph{the normal form}
\begin{eqnarray}\label{eq:NormForm}
  \nonumber \dot{\mathbf{R}(t)}&=&\mathbf{V}(t)\,,\\
   \frac{a\,\tau}{2}\dot{\mathbf{V}(t)}&=&(a-1)\mathbf{V}(t)-\frac{a}{6}\,\beta\,\mathbf{V}(t)^2\,\mathbf{V}(t)\,,
  \end{eqnarray}
which is a normal form of the pitchfork bifurcation and is identical with a normal form of the drift-bifurcation~\cite{PurwinsDS2010}. However, since $u_{0,2}=0$, the delay-induced drift to the lowest order occurs \emph{without change of the shape} (compare to Fig.~\ref{FigDNS} for $\tau=0.5$). The stationary drift velocity can be calculated as $V=\pm\frac{1}{\tau}\,\sqrt{\frac{6\,(a-1)}{\beta\,a}}$ and to the lowest order is given by the same relation as in~\cite{TlidiPRL2009}.

To conclude we have shown that a SHE subjected to TDF exhibits complex spatio-temporal dynamics. The existence of labyrinths indicates that in certain delays the DSHE to some extend behaves like a reaction-diffusion system. The mode truncation, neglecting contributions of the strongly damped modes of the continuous spectrum leads to a system of order parameter equations, explicitly describing the behavior of the single LS in the DSHE. Moreover, a normal form of the delay-induced drift-bifurcation is obtained, showing that induced motion to the lowest order arises without change of the shape. In addition, presented analytical results are derived in general form and can be extended to other spatial extended systems.

We thank Andrei G. Vladimirov for fruitful discussions on the topic.

 \bibliographystyle{apsrev4-1}

%
\end{document}